# Cosmology from the Top Down

**By S.W. Hawking, CH CBE FRS**
*Lucasian Professor of Mathematics*

*Department of Applied Mathematics and Theoretical Physics*  *Email: s.w.hawking@damtp.cam.ac.uk*
*University of Cambridge*  *Web: www.hawking.org.uk*
*Cambridge, CB3 0WA*  *Tel: +44 (0) 1223 337843*
*United Kingdom*  *Fax: +44 (0) 1223 766865*

In this talk, I want to put forward a different approach to cosmology, that can address its central question, why is the universe the way it is. Does string theory, or M theory, predict the distinctive features of our universe, like a spatially flat four dimensional expanding universe with small fluctuations, and the standard model of particle physics. Most physicists would rather believe string theory uniquely predicts the universe, than the alternatives. These are that the initial state of the universe, is prescribed by an outside agency,, code named God. Or that there are many universes, and our universe is picked out by the anthropic principle.

The usual approach in physics, could be described as building from the bottom up. That is, one assumes some initial state for a system, and evolves it forward in time with the Hamiltonian, and the Schroedinger equation. This approach is appropriate for lab experiments like particle scattering, where one can prepare the initial state, and measure the final state.. The bottom up approach is more problem in cosmology however, because we do not know what the initial state of the universe was, and we certainly can't try out different initial states, and see what kinds of universe they produce.

Different physicists react to this difficulty in different ways. Some (generally those brought up in the particle physics tradition) just ignore the problem. They feel the task of physics is to predict what happens in the lab, and they are convinced that string theory, or M theory, can do this. All they think remains to be done, is to identify a solution of M theory, a Calabi-Yau or G2 manifold, that will give the standard model, as an effective theory in four dimensions. But they have no idea why the universe should be four dimensional, and have the standard model, with the values of its 40+ parameters that we observe. How can anyone believe that something so messy, is the unique prediction of string theory. It amazes me that people can have such blinkered vision, that they can concentrate just on the final state of the universe, and not ask how and why it got there.

Those physicists that do try to explain the universe from the bottom up, mostly belong to one of two schools, inflationary models, or pre big bang scenarios. In the case of inflation, the idea is that the exponential expansion, obliterates the dependence on the initial conditions, so we wouldn't need to know exactly how the universe began, just that it was inflating. To lose all memory of the initial state, would require an infinite amount of exponential expansion. This leads to the notion of ever lasting or eternal inflation. The original argument for eternal inflation, went as follows. Consider a massive scalar field in a spatially infinite expanding universe. Suppose the field is nearly constant over several horizon regions, on a space like surface. In an infinite universe, there will always be such regions. The scalar field will have quantum fluctuations. In half the region, the fluctuations will increase the field, and in half, they will decrease it. In the half where the field jumps up, the extra energy density will cause the universe to expand faster, than in the half where the field jumps down. After a certain proper time, more than half the region will have the higher value of the field, because the high field regions will expand faster than the low. Thus the volume averaged value of the field will rise. There will always be regions of the universe in which the scalar field is high, so inflation is eternal. The regions in which the scalar field fluctuates downwards, will branch off from the eternally inflating region, and will exit inflation. Because there will be an infinite number of such exiting regions, advocates of eternal inflation get themselves tied in knots, on what a typical observer would see. So even if eternal inflation worked, it would not explain why the universe is the way it is. But in fact, the argument for eternal inflation that I have outlined, has serious flaws. First, it is not gauge invariant. If one takes the time surfaces to be surfaces of constant volume increase, rather than surfaces of constant proper time, the volume averaged scalar field does not increase. Second, it is not consistent. The equation relating the expansion rate to the energy density, is an integral of motion. Neither side of the equation can fluctuate, because energy is conserved. Third, it is not covariant. It is based on a 3+1 split. From a four-dimensional view, eternal inflation can only be de Sitter space with bubbles. The energy momentum tensor of the fluctuations of a single scalar field, is not large enough to support a de Sitter space, except possibly at the Planck scale, where everything breaks down. For these reasons, not gauge invariant, not consistent, and not covariant, I do not believe the usual argument for eternal inflation. However, as I shall explain later, I think the universe may have had an initial de Sitter stage considerably larger than the Planck scale.

I now turn to pre big bang scenarios, which are the main alternative to inflation. I shall take them to include the Ekpyrotic and cyclic models, as well as the older pre big bang scenario. The observations of fluctuations in the microwave background, show that there are correlations on scales larger than the horizon size at decoupling. These correlations could be explained if there had been inflation, because the exponential expansion, would have meant that regions that are now widely separated, were once in causal contact with each other. On the other hand, if there were no inflation, the correlations must have been present at the beginning of the expansion of the universe. Presumably, they arose in a previous contracting phase, and somehow survived the singularity, or brane collision. It is not clear if effects can be transmitted through a singularity, or if they will produce the right signature in the microwave background. But even if the answer to both of these questions is yes, the pre big bang scenarios do not answer the central question of cosmology, why is the universe, the way it is. All the pre big bang scenarios can do, is shift the problem of the initial state from 13 point 7 billion years ago, to the infinite past. But a boundary condition is a boundary condition, even if the boundary is at infinity. The present state of the universe, would depend on the boundary condition in the infinite past. The trouble is, there's no natural boundary condition, like the universe being in its ground state. The universe doesn't have a ground state. It is unstable, and is either expanding or contracting. The lack of a preferred initial state in the infinite past, means that pre big bang scenarios, are no better at explaining the universe, than supposing that someone wound up the clockwork, and set the universe going at the big bang.

The bottom up approach to cosmology, of supposing some initial state, and evolving it forward in time, is basically classical, because it assumes that the universe began in a way that was well defined and unique. But one of the first acts of my research career, was to show with Roger Penrose, that any reasonable classical cosmological solution, has a singularity in the past. This implies that the origin of the universe, was a quantum event. This means that it should be described by the Feynman sum over histories. The universe doesn't have just a single history, but every possible history, whether or not they satisfy the field equations. Some people make a great mystery of the multi universe, or the many worlds interpretation of quantum theory, but to me, these are just different expressions of the Feynman path integral.

One can use the path integral to calculate the quantum amplitudes for observables at the present time. The wave function of the universe, or amplitude for the metric h i j, on a surface, S, of co-dimension one, is given by a path integral over all metrics, g, that have S as a boundary. Normally, one thinks of path integrals as having two boundaries, an initial surface, and a final surface. This would be appropriate in a proper quantum treatment of a pre big bang scenario, like the Ekpyrotic universe. In this case, the initial surface, would be in the infinite past. But there are two big objections to the path integral for the universe, having an initial surface. The first is the G question. What was the initial state of the universe, and why was it like that. As I said earlier, there doesn't seem to be a natural choice for the initial state. It can't be flat space. That would remain flat space.

The second objection is equally fundamental. In most models, the quantum state on the final surface, will be independent of the state on the initial surface. This is because there will be metrics in which the initial surface is in one component, and the final surface in a separate disconnected component. Such metrics will exist in the Euclidean regime. They correspond to the quantum annihilation of one universe, and the quantum creation of another. This would not be possible if there were something that was conserved, that propagated from the initial surface, to the final surface. But the trend in cosmology in recent years, has been to claim that the universe has no conserved quantities. Things like baryon number, are supposed to have been created by grand unified or electro weak theories, and CP violation. So there is no way one can rule out the final surface, from belonging to a different universe to the initial surface. In fact, because there are so many different possible universes, they will dominate, and the final state will be independent of the initial. It will be given by a path integral over all metrics whose only boundary is the final surface. In other words, it is the so called no boundary quantum state.

If one accepts that the no boundary proposal, is the natural prescription for the quantum state of the universe, one is led to a profoundly different view of cosmology, and the relation between cause and effect. One shouldn't follow the history of the universe from the bottom up, because that assumes there's a single history, with a well defined starting point and evolution. Instead, one should trace the histories from the top down, in other words, backwards from the measurement surface, S, at the present time. The histories that contribute to the path integral, don't have an independent existence, but depend on the amplitude that is being measured. As an example of this, consider the apparent dimension of the universe. The usual idea is that spacetime is a four dimensional nearly flat metric, cross a small six or seven dimensional internal manifold. But why aren't there more large dimensions. Thy are any dimensions compactified. There are good reasons to think that life is possible only in four dimensions, but most physicists are very reluctant to appeal to the anthropic principle. They would rather believe that there is some mechanism that causes all but four of the dimensions to compactify spontaneously. Alternatively, maybe all dimensions started small, but for some reason, four dimensions expanded, and the rest did not.

I'm sorry to disappoint these hopes, but I don't think there is a dynamical reason for the universe to appear four dimensional. Instead, the no boundary proposal predicts a quantum amplitude for every number of large spatial dimensions, from 0 to 10. There will be an amplitude for the universe to be eleven dimensional Minkowski space, i e, ten large spatial dimensions. However, the value of this amplitude is of no significance, because we do not live in eleven dimensions . We are not asking for the probabilities of various dimensions for the universe. As long as the amplitude for three large spatial dimensions, is not exactly zero, it doesn't matter how small it is compared to other numbers of dimensions. We live in a universe that appears four dimensional, so we are interested only in amplitudes for surfaces with three large dimensions. This may sound like the anthropic principle argument that the reason we observe the universe to be four dimensional, is that life is possible only in four dimensions. But the argument here is different, because it doesn't depend on whether four dimensions, is the only arena for life. Rather it is that the probability distribution over dimensions is irrelevant, because we have already measured that we are in four dimensions.

The situation with the low energy effective theory of particle interactions, is similar. Many physicists believe that string theory, will uniquely predict the standard model, and the values of its 40 or so parameters. The bottom up picture would be that the universe would begin with some grand unified symmetry, like E8 cross E8.aS the universe expanded and cooled, the symmetry would break to the standard model, maybe through intermediate stages. The hope would be that String theory, would predict the pattern of breaking, the mass, couplings and mixing angles.

Personally, I find it difficult to believe that the standard model, is the unique prediction of fundamental theory. It is so ugly, and the mixing angles etc, seem accidental, rather than part of a grand design.

In string stroke M theory, low energy particle physics is determined by the internal space. It is well known that M theory has solutions with many different internal spaces. If one builds the history of the universe from the bottom up, there is no reason it should end up with the internal space for the standard model. However, if one asks for the amplitude for a space like surface with a given internal space, one is interested only in those histories which end with that internal space. One therefore has to trace the histories from the top down, backwards from the final surface.

One can calculate the amplitude for the internal space of the standard model, on the basis of the no boundary proposal. As with the dimension, it doesn't matter how small this amplitude is, relative to other possibilities. It would be like asking for the amplitude that I am Chinese. I know I am British, even though there are more Chinese. Similarly, we know the low energy theory is the standard model, even though other theories may have a larger amplitude.

Although the relative amplitudes for radically different geometries, don't matter, those for neighbouring geometries, are important. For example, the fluctuations in the microwave background, correspond to differences in the amplitudes for space like surfaces, that are small perturbations of flat 3 space, cross the internal space. It is a robust prediction of inflation, that the fluctuations are gowsyan, and nearly scale independent. This has been confirmed by the recent observations by the map satellite. However, the predicted amplitude, is model dependent.

The parameters of the standard model, will be determined by the moduli of the internal space. Because they are moduli at the classical level, their amplitudes will have a fairly flat distribution. This means that M theory, can not predict the parameters of the standard model. Obviously, the values of the parameters we measure, must be compatible with the development of life. I hesitate to say, with intelligent life. But within the anthropically allowed range, the parameters can have any values. So much for string theory, predicting the fine structure constant. However, although the theory can not predict the value of the fine structure constant, it will predict it should have spatial variations, like the microwave background. This would be an observational test, of our ideas of M theory compactification.

How can one get a non zero amplitude for the present state of the universe, if as I claim, the metrics in the path integral, have no boundary apart from the surface at the present time. I can't claim to have the definitive answer, but one possibility would be if the four dimensional part of the metric, went back to a de Sitter phase. Such a scenario is realized in trace anomaly driven inflation, for example. In the Lorentzian regime, the de Sitter phase would extend back into the infinite past. It would represent a universe that contracted to a minimum radius, and then expanded again. But as we know, Lorentzian de Sitter can be closed off in the past by half the four sphere. One can interpret this in the bottom up picture, as the spontaneous creation of an inflating universe from nothing. Some pre big bang or Ekpyrotic scenarios, involving collapsing and expanding universes, can probably be formulated in no boundary terms, with an orbifold point. However, this would remove the scale free perturbations which, it is claimed, develop during the collapse, and carry on into the expansion. So again it is a no no for pre big bang and Ekpyrotic universes.

In conclusion, the bottom up approach to cosmology, would be appropriate, if one knew that the universe was set going in a particular way, in either the finite, or infinite past. However, in the absence of such knowledge, it is better to work from the top down, by tracing backwards from the final surface, the histories that contribute to the path integral. This means that the histories of the universe, depend on what is being measured, contrary to the usual idea that the universe has an objective, observer independent, history. The Feynman path integral allows every possible history for the universe, and the observation, selects out the sub class of histories that have the property that is being observed. There are histories in which the universe eternally inflates, or is eleven dimensional, but they do not contribute to the amplitudes we measure. I would call this the selection principle, rather than the anthropic principle because it doesn't depend on intelligent life. Life may after all be possible in eleven dimensions, but we know we live in four.

The results are disappointing for those who hoped that the ultimate theory, would predict every day physics. We can not predict discrete features like the number of large dimensions, or the gauge symmetry of the low energy theory. Rather we use them to select which histories contribute to the path integral. The situation is better with continuous quantities, like the temperature of the cosmic microwave background, or the parameters of the standard model. We can not measure their probability distributions, because we have only one value for each quantity. We can't tell whether the universe was likely to have the values we observe, or whether it was just a lucky chance. However, it is note worthy that the parameters we measure seem to lie in the interior of the anthropically allowed range, rather than at one end. This suggests that the probability distribution is fairly flat, not like the exponential dependence on the density parameter, omega, in the open inflation that Neil Turok and I proposed. In that model, omega would have had the minimum value required to form a single galaxy, which is all that is anthropically necessary. All the other galaxies which we see, are superfluous.

Although the theory can not predict the average values of these quantities, it will predict that there will be spatial variations, like the fluctuations in the microwave background. However the size of these variations, will probably depend on moduli or parameters that we can't predict. So even when we understand the ultimate theory, it won't tell us much about how the universe began. It can not predict the dimension of spacetime, the gauge group, or other parameters of the low energy effective theory. On the other hand, the theory will predict that the total energy density, will be exactly the critical density, though it won't determine how this energy is divided between conventional matter, and a cosmological constant, or quintessence. The theory will also predict a nearly scale free spectrum of fluctuations. But it won't determine the amplitude. So to come back to the question with which I began this talk. Does string theory predict the state of the universe. The answer is that it does not. It allows a vast landscape of possible universes, in which we occupy an anthropically permitted location. But I feel we could have selected a better neighbourhood.